%%%%%%%%%%%%%%% READ THIS %%%%%%%%%%%%%%%% READ THIS %%%%%%%%%%%%%%%%%
%%%%%%%%%%%%%%%%%%%%%%%%%%%%%%%%%%%%%%%%%%%%%%%%%%%%%%%%%%%%%%%%%%%%%%%%%%%
% This paper has figures appended in  a second part as a uuencoded
% compressed tar file with instructions for unpacking.  They will be
%automatically included in the text if you have a functioning epsf.tex.
% If you don't have that macro package (available from hep-th), or don't
% have the figure files, COMMENT OUT THE FOLLOWING LINE:
\input epsf
% If you do not already have epsf.tex (it comes with the dvips driver),
% you can print out the postscript files separately.
% WARNING: there is more than one version of epsf.tex dated 18 Jul 1990
% some figures will produce errors unless you have the most recent
% version. note that the version of epsf.tex that comes with the
% NeXTstep 2.0/2.1 distribution is not an up-to-date version. you should-
% get epsf.tex from hep-th if your dvips is up-to-date but your epsf.tex
% is not.
%
%%%%%%%%%%%%%%%%%%%%%%%%%%%%%%%%%%%%%%%%%%%%%%%%%%%%%%%%%%%%%%%%%%%%%%%%%%%
\input harvmac

%%%%%%%%%%%%%%%%%%
%
%  This inputs the macro package epsf.tex
%
\ifx\epsfbox\UnDeFiNeD\message{(NO epsf.tex, FIGURES WILL BE IGNORED)}
\def\figin#1{\vskip2in}% blank space instead
\else\message{(FIGURES WILL BE INCLUDED)}\def\figin#1{#1}\fi
\def\ifig#1#2#3{\xdef#1{fig.~\the\figno}
\goodbreak\midinsert\figin{\centerline{#3}}%
\smallskip\centerline{\vbox{\baselineskip12pt
\advance\hsize by -1truein\noindent\footnotefont{\bf Fig.~\the\figno:}
#2}}
\bigskip\endinsert\global\advance\figno by1}

\def\ifigure#1#2#3#4{
\midinsert
\vbox to #4truein{\ifx\figflag\figI
\vfil\centerline{\epsfysize=#4truein\epsfbox{#3}}\fi}
\narrower\narrower\noindent{\footnotefont
{\bf #1:}  #2\par}
\endinsert
}
\def\smallfrac#1#2{\hbox{${{#1}\over {#2}}$}}

\Title{\vbox{\baselineskip12pt\hbox{Univ.Roma I, 1146/96}
		\hbox{cond-mat/9606033}}}
{\vbox{\centerline{Site Disordered Spin Systems}
	\vskip2pt\centerline{in the}
	\vskip2pt\centerline{Gaussian Variational Approximation}}}

\centerline{David~S.~Dean\footnote{$^{(a)}$}
  {\tt(dsd@chimera.roma1.infn.it)} and
David~Lancaster\footnote{$^{(b)}$}{\tt(djl@liocorno.roma1.infn.it)}}

\bigskip
\centerline{Dipartimento di Fisica and INFN,}
\centerline{Universit\`a di Roma I {\it La Sapienza},}
\centerline{Piazza A.~Moro 2, 00185 Roma.}

\vskip .3in
Abstract: 
We define a replica field theory describing finite dimensional site 
disordered spin systems by introducing the notion of grand canonical 
disorder, where the number of spins in the system is random but quenched.
A general analysis of this field theory is made using the Gaussian
variational or Hartree Fock method, and illustrated with
several specific examples.
Irrespective of the form of interaction between the spins
this approximation predicts a spin glass phase.
We discuss the replica symmetric phase at length, explicitly
identifying the correlator that diverges at the spin glass transition.
We also discuss the form of continuous replica symmetry breaking found
just below the transition.
Finally we show how an analysis of ferromagnetic ordering
indicates a breakdown of the approximation.

%\draft
\Date{3/96} %replace this line by \draft  for preliminary versions
	     %or specify \draftmode at some point

%%%%%%%%%%%%%%%%%%%%%%%%%%%%%%%%%%%%%%%%%%%

\lref\SK{D.~Sherrington and S.~Kirkpatrick, Phys.Rev.Lett., {\bf 35},
 1972 (1975)\semi Phys.Rev.  {\bf B 17}, 4384 (1977).} 
\lref\She{D.~Sherrington, Phys.Rev. {\bf B 22}, 5553 (1980).} 
\lref\McSh{I.R.~McLenaghan and D.~Sherrington, J.Phys. {\bf C 17}, 1531 
(1984).} 
\lref\EdAn{S.~F.~Edwards and P.~W.~Anderson, J.Phys. {\bf F 5}, 965 (1975).}
\lref\Nie{Th.~M.~Nieuwenhuizen, Europhys.Lett., {\bf 24}, 797 (1993).}
\lref\MePaVi{M. M\'ezard, G. Parisi and Virasoro M. A., {\it Spin Glass 
 Theory and Beyond}, Lecture Notes in Physics Vol. 9, World Scientific,
 Singapore, (1987).}
\lref\FiHu{D.S.~Fisher and D.A.~Huse, Phys.Rev.Lett.  {\bf 56}, 1601 (1986)
	\semi
 	Phys.Rev.{\bf B 38}, 373 (1988).}
\lref\Remi{M.~M\'ezard and A.~P.~Young, Europhys.Lett. {\bf 18}, 653 (1992)
	\semi
	M.~M\'ezard and R.~Monasson, Phys.Rev. {\bf B 50}, 7199 (1994).}
\lref\DHSS{V. Dotsenko, A.B. Harris, D. Sherrington and R.B. Stinchcombe,
  J.Phys. {\bf A 28}, 3093 (1995).}
\lref\LDGi{P.~Le Dousal and T.~Giamarchi, Phys.Rev.Lett. {\bf 74}, 606
  (1995)\semi
	J.~Kierfeld,  J.Phys. {\bf I} France 5, 379 (1995).}  
\lref\FiHe{K.~H.~Fischer and J.~A.~Hertz, {\it Spin Glasses}, Cambridge
 University Press, Chapter 11, (1991).}
\lref\DeLa{D.S. Dean and D.J. Lancaster, cond-mat 9602070.}
\lref\DeLaRG{D.S. Dean and D.J. Lancaster, in preparation.}
\lref\AlTh{J.R.L. de Almeida and D.J. Thouless, J.Phys.  {\bf A 11} 983
(1978).}
\lref\ThankG{We thank G.~Parisi for bringing this argument to our 
	attention.}
\lref\largeN{D.S. Dean and D.J. Lancaster, in preparation.}
\lref\GandM{M. M\'ezard and G. Parisi, J. Phys. I France {\bf 1}, 809 
(1991.)}
\lref\PytteR{E. Pytte and J. Rudnick, Phys.Rev. {\bf B 19}, 3603 (1979).}
\lref\Pa{G. Parisi, J. Phys. {\bf A 13}, 1101 (1980).}
\lref\Kondor{I. Kondor, C. De Dominicis and T. Temesv\'ari, 
 J.Phys. {\bf A 27}, 7569 (1994).}
\lref\Feynman{R.~P.~Feynman, {\it Variational Calculations in Quantum
  Field Theory}, Wangerooge Proc., World Scientific, (1987)}
\lref\Shaknovich{E.~I.~Shakhnovich and A.~M.~Gutin, 
	J.Phys. {\bf A 22}, 1647 (1989).}
\lref\ParisiBook{G.~Parisi, {\it Statistical Field Theory}, Addison-Wesley,
	(1988).}
\lref\ThAlKo{D.J. Thouless, J.R.L. de Almeida and J.~M.~Kosterlitz,
	J.Phys. {\bf C 13}, 3271, (1980).}
\lref\PaDu{G.~Parisi, J. Phys. {\bf A 13}, L115 (1980)
	 \semi B.~Duplantier,  J.Phys.  {\bf A 14}, 283 (1981).}
\lref\Remib{R.~Monasson, Phys.Rev.Lett. {\bf 75}, 2847 (1995).}
\lref\Giulia{G.~Iori, Private Communication.}
\lref\ViBr{L.~Viana and A.J.~Bray, J.Phys. {\bf C 18}, 3037 (1985)
	\semi C.~De~Dominicis and P.~Mottishaw, J.Phys. {\bf A 20}, L1267
	(1987).}

%%%%%%%%%%%%%%%%%%%%%%%%%%%%%%%%%%%%%%%%%%%

\newsec{Introduction}

We consider site disordered models in finite dimensions
with the following Hamiltonian:
\eqn\HamiltonianIntro{
H = -{1\over 2} \sum_{ij} J(r_i - r_j)S_iS_j
}
in which $N$ Ising spins are fixed at random points, $r_i$, 
and are subject to a deterministic potential $J(r)$. 
For example, a positive $J(r)$ decaying
with distance describes a dilute ferromagnetic system.
Antiferromagnetic systems can also be treated,
and because there is no lattice in this picture there will
be no antiferromagnetic ordering.
We also have in mind oscillatory RKKY--type interactions 
which cause frustration, as in the antiferromagnetic case.
The physics we therefore address is related to the two types of
ordering that can occur; ferromagnetic and spin glass, and to the
relation between them.
We shall discover that the approximation we use to solve the model
is not completely reliable for the ferromagnetic system,
nevertheless we present the general development for
ferromagnetic interactions indicating where sign differences arise
for the antiferromagnetic case.
The Hamiltonian \HamiltonianIntro\ could also be interpreted as 
an infinite range Sherrington Kirkpatrick model \SK\ in which the 
bond strengths are correlated and have been chosen from a
distribution very different from those we usually consider.

This Hamiltonian can describe experimental systems rather well.
However, analytic studies of disordered spin systems
are usually based on lattice models in which the bonds take random values,
such as the Edward Anderson Hamiltonian \EdAn.
Analytic work based on \HamiltonianIntro\ 
has been hampered by the lack of a suitable field theoretic model 
(however a lattice based formulation has been proposed \Nie).
By considering a situation in which 
the number of spins in the system is random but quenched
we are able to write a replica field theory for these
site-disordered systems \DeLa.
This field theory seems to be simpler than many of those coming from 
bond-disordered and diluted lattice models and should be accessible
to many standard analytical techniques.

In this paper, to gain an overall picture of the model, 
we use the technique variously known as Gaussian variational, 
Hartree Fock, Random Phase approximation and other names.
It is necessary to use a method more sophisticated than mean field theory 
because the field theory that describes the model is expressed in terms 
of the magnetic variables and to understand the spin glass physics 
one has to look at composite operators. Mean field theory alone 
misses this spin glass physics, whereas the Gaussian variational 
method indicates that a spin glass transition is ubiquitous.
We find that using the Gaussian variational approximation the model can
be solved in considerable detail, analytically in the high temperature
region, and with the help of numerics at low temperature.
The general picture that emerges is in accord with physical
expectations, however there are some points that throw the reliability 
of the approximation into doubt. This is not surprising since 
although the approximation is expected to be good in the case
of Heisenberg spins with many components, here we apply it
to one--component Ising spins. The infinite component spin
case is discussed in a companion paper \largeN. 
As we shall show, the problems with the approximation appear most
clearly in two issues. Firstly, we shall find that for a purely
ferromagnetic interaction a spin glass transition is predicted
to occur at slightly higher temperature than the ferromagnetic 
transition and we present a proof of the impossibility of such a 
situation. Secondly, the approximation yields results that are
not as dependent on the range of interaction or dimension as
one would expect. This should act as a caution given the
rather widespread application of the method in the literature.
For this reason and also because a thorough understanding of this
simplest non-trivial approach is a prerequisite for further advances,
we feel that this study is worthwhile despite the deficiencies
of the approximation.

One of the interesting aspects of this work is that
we are able to investigate the spin glass transition in 
a finite dimensional model.
In the formalism we present, replicas enter the theory 
at an early stage and the spin glass ordering is
intimately connected with the symmetry in replica space.
The role of replica symmetry breaking (RSB) is of great interest 
in spin glasses, where the correct description of three dimensional
materials is still controversial. Although the RSB in the mean field 
theory for spin glasses is now well understood \MePaVi\ and related
to the proliferation of pure states of the system, 
in finite dimensions the picture is less clear.
Alternative qualitative approaches based on droplets \FiHu\ view the
spin glass phase as a disguised ferromagnetic phase with only 
two underlying fundamental states.
The role of RSB in systems which undergo a ferromagnetic phase
transition is maybe even less clear. We might mention the random
field Ising model \Remi, 
dilute ferromagnets \DHSS\ and the whole issue of
renormalisation flow in the presence of RSB \LDGi.
Experimentally, the problem is reflected in difficulty with
the so-called reentrant transitions \FiHe.
Part of the theoretical difficulty in these cases lies
in the lack of a simple model in which the ferromagnetic transition
can be explicitly analysed in the presence of RSB.
All these issues provide strong motivation to study 
the model \HamiltonianIntro.

Some of this work has already been briefly reported in \DeLa,
here we attempt a more pedagogic approach, discussing
the basic issues more throughly and giving detailed examples.
The paper is organised as follows, Section 2 explains ``grand
canonical disorder'' constructs the field theory and 
discusses some of the measurement issues.
In section 3 we perform the Gaussian variational analysis of the theory
to derive the variational equations.
These equations are discussed for a general potential 
in sections 4 and 5 which concern the replica symmetric (RS)
and RSB cases respectively.
In the remaining sections we illustrate our
results with reference to some particularly simple interactions,
namely Yukawa and RKKY-type.
A discussion of problems with this approach 
along with some speculation, appears in the conclusion. 
Two appendices give details of the solution of the four
index correlator and of the stability analysis.

%%%%%%%%%%%%%%%%%%%%%%%%%%%%%%%%%%%%

\newsec{Grand Canonical Disorder}

Firstly consider a model where the number of spins $N$ is fixed: 
$N$ spins $S_i$ are placed randomly at positions $r_i$ 
uniformly throughout a volume $V$.
This type of disorder we  refer to as 
canonical disorder, as the number of particles is the same for each
realization of the disorder. The spins interact via a pairwise potential $J$
depending only on the distance 
between the spins. 
The Hamiltonian is the one written down in the introduction,
\eqn\Hamiltonian{
H = -{1\over 2} \sum_{ij} J(r_i - r_j)S_iS_j
}
We shall proceed in the derivation assuming that $J(r)$ is positive,
thus giving rise to ferromagnetic interactions. Later, we shall also discuss
purely antiferromagnetic interactions, and at stages in the development
will point out the sign changes necessary for a negative potential.
A Hubbard-Stratonovich transformation expresses the partition function as 
\eqn\PartitionN{
Z_N = 
\sum_{S_i}\int {\cal D} \phi 
[det J\beta]^{-{1\over 2}}
{\rm exp} \left( -{1\over 2 \beta}\int\!\int  
\phi(r)J^{-1}(r-r')\phi(r')\ drdr'
+\sum_i^N\phi(r_i)S_i \right)
}
Employing replicas,  
we average out the site-disorder by integrating over the
positions  $r_i$ using the flat measure: 
${1\over V^N}\int_V \prod dr_i $. 
\eqn\AvPartitionNn{\eqalign{
{\overline Z^n} &=
\int {\cal D} \phi_a 
[det J\beta]^{-{n\over 2}}
{\rm exp} \biggl( -{1\over 2 \beta}\int\!\int \sum_a 
\phi_a(r)J^{-1}(r-r')\phi_a(r')\ drdr'\cr
&+N\log {1\over V}\int {\rm Tr}\ \exp 
\bigl(\sum_a\phi_a(r)S_a\bigr)\ dr \biggr)\cr}
}
Where we have introduced the trace over single site spins $S_a$ 
as convenient notation rather
than write explicit $cosh \phi_a$ factors.
A field theoretic analysis of the above theory is complicated by the
presence of the $\log$ term in the action. 
We overcome this difficulty by making
a physically desirable modification to the definition of the disorder. In 
general one might expect the system to have been taken 
from a much larger system with a mean concentration of spins 
per unit volume, $\rho$. 
A suitably large  
subsystem of volume $V$ will thus contain a number of spins
$N$ which is random and Poisson distributed:
$P(N) = e^{-\rho V}  {(\rho V)^N\over N!}$.
This distribution must be used to weight the averaged free energy,
\eqn\Poisson{\eqalign{
-\beta F_\Xi &= -\beta \sum_N P(N) {\overline F_N}
= \sum_N P(N) {\overline {\log Z_N}}\cr
&= \lim_{n\to0}\sum_N P(N) \left({{\overline Z_N^n} -1\over n}\right)
= \lim_{n\to0} \left({\Xi^n -1\over n}\right)\cr}
}
so we are led to define the partition function,
$ {\Xi^n} = \sum_N P(N){\overline Z_N^n}$.
By analogy with the statistical mechanics of pure systems, 
we shall call  this type of disorder ``grand canonical disorder''. 

The theory is defined by the partition function for grand canonical disorder:
\eqn\GCPartition{\eqalign{
\Xi^n 
&= e^{-\rho V}
\int {\cal D} \phi_a 
[det J\beta]^{-{n\over 2}}
{\rm exp} \biggl( -{1\over 2 \beta}\int\!\int \sum_a 
\phi_a(r)J^{-1}(r-r')\phi_a(r')\ drdr'\cr
&+ \rho\int {\rm Tr}\ \exp \bigl(\sum_a\phi_a(r)S_a\bigr) \ dr \biggr)\cr}
}
Expanding the trace one sees that the leading term mixing replicas 
corresponds to a random temperature or random mass, 
familiar from bond disordered approaches, and that depending
on the choice of interaction one might 
expect similar renormalisation group results \refs{\DeLaRG, \DHSS}.
The simplicity of this form is evident, but the physical content is not 
immediately obvious so we now turn to a discussion of this point.

%%%%%%%%%%%%%%%%%%%%%%%%%%%%%%%%%%%%%%%%%%%%

\subsec{Physical Operators}

In order to relate this theory to measurable
quantities we must understand the type of excitations occurring
in the theory and identify them with physical operators.
This is not a simple task since the bound states of \GCPartition\
are certainly not apparent at first glance. 

We start by returning to
the original formulation of the model 
and considering the operator most closely associated with the 
field $\phi_a$ appearing in the theory.
This is the spin density operator,
\eqn\SpinDensity{
M_a(r)=\sum_i \delta(r-r_i)S_i^a}
which appears in the replicated but
unaveraged action as a source for $\phi_a$.
\eqn\PartitionNn{
Z^n =
\int {\cal D} \phi_a 
[det J\beta]^{-{n\over 2}}
{\rm exp} \left( -{1\over 2 \beta}\int\!\int \sum_a 
\phi_aJ^{-1}\phi_a\ drdr'
+\int \sum_a M_a\phi_a\ dr \right)
}
The equations of motion (or Ward identities) at this 
unaveraged level provide relations between expectation values
of $M_a$'s and $\phi_a$'s. 
The equations of motion constitute a powerful tool and 
continue to hold for the fully averaged theory.
For example the simplest relation is
$\langle M_a \rangle = \langle\phi_a\rangle /\beta\tilde J(0)$.
To make a connection with measurements performed on
a particular sample with given disorder, we must look
at self averaging quantities. One such quantity is the
magnetisation density $M$, and below we show how it is related 
to the averaged theory, the replicated theory and finally
to the expectation values of fields $\phi_a$.
\eqn\Magnetisation{ 
M \ {\buildrel{large\ V}\over\longrightarrow}  \ 
{1\over V}\int \!dr \overline{\langle M(r)\rangle} 
= \rho \Bigl[ \langle S \rangle \Bigr]_{av}
= \lim_{n\to 0}{1\over nV}\sum_a\int \!dr \langle M_a(r) \rangle
={\langle\phi\rangle\over\beta\tilde J(0)}.}
The over-line denotes the average over the full grand canonical
disorder. In the third expression we have inserted the
definition of the magnetisation density operator to
show the relation to the spins, and in this case
the square brackets indicate the average over site disorder alone.
In the last term we have assumed that $ \langle\phi_a\rangle$ is independent
of the replica index.

Similar arguments can be applied to the correlation function
of magnetisation density operators.
Firstly, 
$\langle M_a(r)M_b(r')\rangle $, is related to the
field correlator 
$\langle \phi_a(r)\phi_b(r') \rangle$, 
by,
\eqn\MagCorrelator{
 \langle \tilde M_a(k) \tilde M_b(-k)\rangle
= {\langle \tilde\phi_a(k)\tilde\phi_b(-k) \rangle
\over\beta^2\tilde J^2(k)}
-{\delta^{ab}\over \beta \tilde J(k)} \ .}
Where we have assumed spatial translation invariance.
In an experiment, the neutron elastic scattering cross section
is proportional to the following correlator,
\eqn\MagPhysical{
{\overline {\langle \tilde M(k) \tilde M(-k)\rangle}}
=\rho^2 \Bigl[ \langle \tilde S(k)\tilde S(-k) \rangle \Bigr]_{av}
=\lim_{n\to 0}{1\over n}\sum_{a}
\langle \tilde M_a(k) \tilde M_a(-k)\rangle
}
The connected and disconnected parts of this correlation
function can also be determined separately.

The spin density operator we have considered so far
is a simple generalisation within
the grand canonical context of the familiar spin operator.
The expectation value, $M$ in \Magnetisation,  is the order
parameter for ferromagnetism, but $M_a(r)$ is
certainly not the operator sensitive to spin glass ordering.
To probe this aspect of the physics 
it is natural to consider another operator:
\eqn\Qab{
q_{ab}(r) = \sum_i \delta (r -r_i) S_i^a S_i^b.}
This operator is familiar from spin glass physics,
its expectation value is a order parameter for the transition
and its correlator is
related to the non-linear susceptibility.
In fact, bond disordered models such as the Edwards Anderson
model lead to field theories \PytteR\ in which the basic field variable
is $q_{ab}(r)$ itself. These models immediately lead to
non-trivial mean field theories describing the spin glass
order parameter.
The  theory \GCPartition, on the other hand, contains only the simple
fields $\phi_a$ so the spin glass physics is hidden in the
composite operators \Qab\ which do not appear at mean field
level and necessarily require a deeper understanding of
the field theory.
In general we cannot provide such a simple and general relation 
between expectation values of $q_{ab}(r)$ and of the 
fields $\phi_a$ as we were able to for
the spin densities. However, within the context of the variational
approximation, we will see that using linear response we can
derive expressions for $\langle q_{ab}(r)\rangle$ 
and equations obeyed by the correlators 
$\langle \tilde q_{ab}(k) \tilde q_{cd}(-k) \rangle$.
The connection between $\langle q_{ab}(r)\rangle$ and the
usual physical order parameter of Edwards and Anderson
is clear from the relation 
$\langle q_{ab}(r)\rangle = \rho \bigl[ \langle S\rangle^2\bigr]_{av}$.
Although physical observables do not directly measure the
correlators, these functions form a central part of our understanding
in finite dimensions of 
both bond-disordered spin glasses and the site disordered
systems considered here.
We shall concentrate our attention on the connected 
part of the spin correlator which is related to the $q_{ab}$
correlators as follows:
\eqn\QSpin{
\rho^2 \Bigl[ \langle \tilde S(k)\tilde S(-k) \rangle^2_{con} \Bigr]_{av}
=\lim_{n\to 0}{1\over n(n-1)}\sum_{a\ne b}
\Bigl(\langle \tilde q_{ab}(k) \tilde q_{ab}(-k) \rangle
-\langle \tilde q_{aa}(k) \tilde q_{bb}(-k) \rangle\Bigr) \ .
}
More detailed information about the theory could be deduced by studying
higher moments of the spin probability distribution. To do this,
operators involving more spins can be introduced in a similar way.

%%%%%%%%%%%%%%%%%%%%%%%%%%%%%%%%%%%%

\newsec{Variational Method}

As we have already argued, mean field theory does not describe the
interesting physics of this model and one must consider some improvement.
In this section we analyse the theory using the Gaussian variational method
which will allow us to substantially develop the theory 
without making an explicit choice for the form of the interaction $J(r)$.
The Gaussian variational equations
are truncations of the full Schwinger Dyson equations and become exact in 
the limit of many spin components (such an $m$-component
theory is treated in a separate publication \largeN).
We find it convenient to phrase the development in terms 
of a variational method which allows us to calculate thermodynamic
quantities simply, but from the outset, we would like to acknowledge
the difficulties of variational methods for field theory \Feynman. 
In the context of disordered systems, this method has had success in
calculating exponents for random manifolds \GandM,
and earlier in the problem of random heteropolymers \Shaknovich.
On the other hand, it has been less successful in situations where the RSB
occurs in a one step pattern. One should also 
bear in mind that important effects may occur at higher orders in $1/m$
and indeed we shall discover that the approximation does not 
obey certain requirements expected on general grounds for purely
ferromagnetic interactions.

From the variational point of view, one selects a trial Hamiltonian, $H_t$,
and the method may be motivated as follows:
\eqn\Variational{\eqalign{
e^{-n\beta F_\Xi} &= \Xi^n = e^{-\rho V}\int {\cal D} \phi e^{-(H-H_t) -H_t}
= \langle e^{-(H-H_t)} \rangle_{t} \ e^{-F_t}\ e^{-\rho V}\cr
&>  e^{-\langle(H-H_t)\rangle_{t}} \ e^{-F_t}\ e^{-\rho V}\ .\cr}}
So the variational free energy,
$n\beta F_{var} = F_{trial} + \langle H - H_{trial}\rangle_{trial}+\rho V$,
provides a bound to the true free energy: $F_\Xi < F_{var}$.
When the replica limit is taken, this bound is no longer rigorous,
nevertheless the expression for the variational free energy
is still valid. 
The Gaussian variational method simply consists in 
making a Gaussian trial, that is, choosing
a trial Hamiltonian that is quadratic.
Mean field theory can also be viewed as variational, in which case
it corresponds to a linear trial Hamiltonian, and in this sense 
the Gaussian variational method is
a simple generalisation of mean field theory.

We allow the possibility of ferromagnetic order and make
the following Gaussian anzatz 
(in which we have assumed translational invariance),
\eqn\Htrial{
H_{trial}={1\over 2}\int \sum_{ab} 
(\phi_a(r)-\bar\phi_a)G_{ab}^{-1}(r-r')(\phi_b(r')-\bar\phi_b)\ drdr'\ .
}
The variational parameters, $\bar\phi_a$ and $G_{ab}(r)$ are simply
related to $\phi$ expectation values: 
$\langle \phi_a \rangle = \bar\phi_a$ and 
$\langle \phi_a(r)\phi_b(r') \rangle = G_{ab}(r-r')$. 
This gives rise to a variational free energy of the following form,
\eqn\Fvar{\eqalign{
n\beta F_{var}&=
-{1\over 2} {\rm Tr} \log G_{ab}
+{1\over 2 \beta}\int\!\int \sum_a 
\left( \bar\phi_aJ^{-1}(r-r')\bar\phi_a 
+G_{aa}(r-r')J^{-1}(r-r') \right)\ drdr'\cr
&\quad -\rho{\rm Tr}\int {\rm exp}\left(\sum_a \bar\phi_aS_a
+{1\over 2}\sum_{ab} G_{ab}(r,r)S_aS_b \right)\ dr 
+{n\over 2}{\rm Tr}\log (\beta J - 1) + \rho V .\cr
&=
-{V\over 2} \int {d^dk\over (2\pi)^d}
\sum_a\left[ \log \tilde G(k)\right]^{aa}
+{V\over 2 \beta}\int {d^dk\over (2\pi)^d}
\sum_a \tilde G_{aa}(k)\tilde J^{-1}(k) \cr
&\quad +{V\over 2 \beta}\sum_a \bar\phi_a^2\tilde J^{-1}(0) 
-\rho V (\Omega - 1) 
+{nV\over 2} \int {d^dk\over (2\pi)^d} \left( \log \beta J - 1\right) \ .\cr}
}
We have kept all the constant terms and have defined,
\eqn\Omegadef{
\Omega = {\rm Tr}\ e^{H'}
=\sum_{S_a = \pm 1}\ {\rm exp}\left(\sum_a \bar\phi_aS_a
+{1\over 2}\sum_{ab} G_{ab}(0)S_aS_b \right).
}
The variational equations follow immediately by varying with
respect to $\bar\phi_a$ and $G_{ab}$. Their general form is:
\eqn\Varequations{
\eqalign{
\bar\phi_a J^{-1}(0) &= \rho\beta\Omega_{a}
=  \rho\beta{\rm Tr}\ S_a\ e^{H'}\cr
\tilde G_{ab}^{-1} &= {1\over \beta}\delta_{ab}\tilde J^{-1} 
-\rho \Omega_{ab}\cr}}
where we have introduced generalisations of \Omegadef :
\eqn\Omegaabdef{\eqalign{
\Omega_{a} &= {\rm Tr}\, S_a\, e^{H'}
=\sum_{S_a = \pm 1}S_a \, {\rm exp}\left(\sum_a \bar\phi_aS_a
+{1\over 2}\sum_{ab} G_{ab}(0)S_aS_b \right)\cr
\Omega_{ab} &= {\rm Tr}\, S_aS_b\, e^{H'}
=\sum_{S_a = \pm 1} S_aS_b \, {\rm exp}\left(\sum_a \bar\phi_aS_a
+{1\over 2}\sum_{ab} G_{ab}(0)S_aS_b \right)\cr}
}
Further generalisations involving more indices will 
appear in future equations.

The variational equations, \Varequations, 
can be interpreted diagrammatically as
the leading equations of Schwinger Dyson, truncated in 
at the level of the four point function.
As usual in this type of calculation, 
we find that $\Omega_{ab}$ is momentum independent;
in many applications this only leads to mass renormalisation,
but due to the replica limit non-trivial effects can occur \GandM.
There exist some similarities with the formalism of \Nie.
The next few sections will be devoted to solving these equations
assuming various forms for the replica structure.

The above derivation is for a ferromagnetic potential,
in general the interaction potential is not positive definite and to deal
with it correctly one must split up the $k$ space range accordingly.
However there is a simple case, that  of
a purely negative or antiferromagnetic interaction.
There will be no magnetisation, $\bar\phi = 0$ in this case, and
at the level of the free energy, the effect is a sign change of 
$G_{ab}$ in the definition of $\Omega$ \Omegadef.
This leads to a reversed sign in the second variational equation
and to a similar redefinition of $\Omega_{ab}$.

Using standard thermodynamic relationships 
we can also determine the entropy and internal energy
from the free energy \Fvar,
\eqn\Entropy{
\eqalign{
{n S\over V} &=
{1\over 2} \int {d^dk\over (2\pi)^d}
\sum_a\left[ \log \tilde G(k)\right]^{aa}
-{1\over \beta}\int {d^dk\over (2\pi)^d}
\sum_a \tilde G_{aa}(k)\tilde J^{-1}(k) \cr
&\quad -{1\over  \beta}\sum_a \bar\phi_a^2\tilde J^{-1}(0) 
+\rho (\Omega - 1) 
+{n\over 2} \int {d^dk\over (2\pi)^d} \left(2- \log \beta J\right) \cr
{n U\over V} &=
-{1\over 2\beta^2}\int {d^dk\over (2\pi)^d}
\sum_a \tilde G_{aa}(k)\tilde J^{-1}(k) 
 -{1\over 2 \beta^2}\sum_a \bar\phi_a^2\tilde J^{-1}(0) 
+{n\over 2\beta}\int {d^dk\over (2\pi)^d}\ .\cr}
}
At high temperature, in the RS phase with zero magnetisation,
these formulae simplify considerably.
We find, $S = \rho V \log 2$ as expected for Ising spins,
and $U = -\rho J(0) /2$ because in this limit the only spin 
correlation is from the same spin term of \Hamiltonian.

%%%%%%%%%%%%%%%%%%%%%%%%%%%%%%%%%%%%%%%%%%%%

\subsec{Physical Operators in the Gaussian Approximation}

In the context of this approximation
we can calculate expectation values
of products of the physical operators identified in 
section 2.1. In general for variational theories
these quantities should be determined using linear response
as this leads, in the variational sense, to a smaller error than a
direct evaluation \ParisiBook.
At the level of $\overline Z^n$ \PartitionNn, 
we introduce suitable sources for the operators of interest,
follow through the analysis to obtain the generalised
variational free energy, and then take appropriate derivatives
before setting the sources to zero.
Sources for the operators $M_a(r)$ and $q_{ab}(r)$, only modify the
free energy \Fvar\ through $\Omega$, which 
becomes,
\eqn\OmegaFDT{
\Omega (r) = {\rm Tr}\ e^{H'(r)}
={\rm Tr}\ {\rm exp}\left(\sum_a \bigl(\bar\phi_a(r) + \beta h_a(r)\bigr)S_a
+{1\over 2}\sum_{ab} \bigl(G_{ab}(r,r) + \beta j_{ab}(r)S_aS_b \bigr) \right).
}
where $h_a(r)$ and $j_{ab}(r)$ are the respective sources
for $M_a(r)$ and $q_{ab}(r)$. Translational
invariance is lost until the sources are set to zero at the
end of the calculation.

To illustrate the method, consider the spin density operator $M_a(r)$.
The magnetisation is defined as:
\eqn\MagnetisationFDT{
M_a(r) = -{1\over \beta} {\delta F_{var}\over \delta h_a(r)} = \rho\Omega_a(r)
 \ .}
Using the variational equation of motion for $\bar\phi$ \Varequations\
we recover the equation 
$M_a(r) ={1\over \beta} \tilde J^{-1}  \bar\phi_a$
that we used in \Magnetisation. In fact, 
this method merely reproduces the equations of motion we had earlier
for expectation values of spin density operators,
and the correlation function is given by \MagCorrelator.

For the spin glass operator, $q_{ab}(r)$, there is no alternative 
method and in this case applying linear response gives:
\eqn\qabFDT{\eqalign{
\langle q_{ab}(r) \rangle 
&= -{1\over \beta} {\delta F_{var}\over \delta j_{ab}(r)} = 
\rho\Omega_{ab}(r)\cr
Q_{abcd}(r - r') =
\langle q_{ab}(r) q_{cd}(r') \rangle
& = -{1\over \beta^2}
{\delta^2 F_{var}\over \delta j_{ab}(r) \delta j_{cd}(r')}
= {\rho\over\beta}{\delta \Omega_{ab}(r)\over \delta j_{cd}(r')}\cr}
}
Using the variational equation of motion, some appropriate inverses
and setting magnetisations to zero we finally obtain the equations:
\eqn\Qequation{\eqalign{
\tilde Q_{abcd}(k)&= \rho \Omega_{abcd}
+{\rho\over 2}\sum_{gh} \tilde \Sigma_{abgh}(k) \tilde Q_{ghcd}(k)\cr
\tilde\Sigma_{abgh}(k)&= \sum_{ef}\Omega_{abef}\int {d^dp\over (2\pi)^d}
\tilde G_{eg}(p)\tilde G_{fh}(k-p)\cr}
}
Where $\Omega_{abcd}$ is a trace of the form \Omegadef\ containing
four spins. 
This equation goes beyond the simple Gaussian
approximation, as is clear from its diagrammatic interpretation.
It may be helpful for intuition to note that a similar diagram
appears in the BCS theory of superconductivity.
One reason for the importance of being able to get explicit
expressions for the spin glass correlators is their use in
showing the breakdown of the approximation for ferromagnetic
interactions in section (4.2).

Equations of this type, involving objects with
four replica indices are familiar in the theory of spin glasses.
In this particular case, note that although
symmetry in $a\leftrightarrow b$ and $c\leftrightarrow d$ is manifest,
the symmetry between pairs $(ab)\leftrightarrow (cd)$ is not. 
Also remember that, in contrast to the SK mean field situation,
diagonal terms with two equal indices do not vanish. 
In appendix A we solve this equation in the RS case but in the 
case of continuous RSB an extension of the methods of \Kondor\
would be needed.

%%%%%%%%%%%%%%%%%%%%%%%%%%%%%%%%%%%%

\newsec{Replica Symmetric Case}

We now specialise to Replica Symmetric (RS) 
solutions of the equations.
Our parameterisation of $G_{ab}$ is: $g_0 + g_1$ on the diagonal
and, $g_1$ elsewhere.
The constant term $\bar\phi_a$ is taken to be replica independent.
In this RS case we may write an explicit expression for $\Omega$ \Omegadef,
which leads to the following form (with ferromagnetic signs) 
for the variational free energy,
\eqn\FvarRS{\eqalign{
\beta {F_{var}\over V}&=
-{1\over 2}\int {d^dk\over (2\pi)^d}
 \left( \log \tilde g_0 + {\tilde g_1\over \tilde g_0} \right)
+ {1\over 2 \beta}\int {d^dk\over (2\pi)^d}
\left(\tilde g_0+\tilde g_1\right)\tilde J^{-1}
+ {1\over 2 \beta} \bar\phi^2 \tilde J^{-1}(0) \cr
&-{\rho\over 2}g_0(0)
-{\rho\over \sqrt{2\pi}}\int d\xi e^{-{\xi^2\over 2}}
\log \left(2 \cosh(\bar\phi+\xi\sqrt{g_1(0)})\right)
+{1\over 2} \int {d^dk\over (2\pi)^d} \left( \log \beta J - 1\right)\cr}
}
from which the RS version of the
variational equations may be read off.

According to the first of equations \qabFDT,
we define the Edward Anderson order parameter $q$ as $\Omega_{ab}$
for $a\ne b$.
This order parameter takes values in $[0,1]$ and is given by:
\eqn\qdefn{
q=\Omega_{a\ne b}={1\over \sqrt{2\pi}}\int d\xi e^{-{\xi^2\over 2}}
{\rm tanh}^2(\bar\phi+\xi\sqrt{g_1(0)})
}
The variational equations for the components of the correlator
are simply solved to give,
\eqn\GsolnRS{
\eqalign{
\tilde g_0(k) &=
{\beta \tilde J(k)\over 1-(1-q)\rho\beta \tilde J(k)}\cr
\tilde g_1(k) &= \rho q\tilde g_0^2(k)
={\rho q\beta^2 \tilde J^2(k)\over 
\left(1-(1-q)\rho\beta \tilde J(k)\right)^2}\cr
}}
The value of the element $g_1$ at zero spatial distance, $g_1(0)$,
appears throughout the equations and can be determined from
the momentum space form above as,
\eqn\GoneRS{
g_1(0)
= \rho q\int {d^dk\over (2\pi)^d}
{\beta^2 \tilde J^2(k)\over \left(1-(1-q)\rho\beta \tilde J(k)\right)^2}
}

The equation for $\bar \phi$ is simply re-expressed in terms of the 
magnetisation $M = \bar\phi/\beta \tilde J(0)$ \Magnetisation.
Together with the expression for $q$, we obtain a pair
of equations defining the two order parameters and which
bear a striking resemblance to the RS mean field equations 
for the Sherrington Kirkpatrick model \SK,
\eqn\MFTRS{
\eqalign{
M
&={\rho\over \sqrt{2\pi}}\int d\xi e^{-{\xi^2\over 2}}
\tanh\left(\beta\tilde J(0)M + \xi\sqrt{g_1(0)}\right)\cr
q
&={1\over \sqrt{2\pi}}\int d\xi e^{-{\xi^2\over 2}}
{\rm tanh}^2\left(\beta\tilde J(0)M + \xi\sqrt{g_1(0)}\right)\cr}
}
In general a numerical solution is necessary, but 
at high temperature and low density a unique 
$q=0$, $M=0$ paramagnetic solution is expected.
Depending on the interaction potential, which may or may
not allow a ferromagnetic state, the solution
leads to two types of critical line corresponding
to the order parameters $M$  of ferromagnetism and to $q$
of spin glass order which divide the phase diagram in
the temperature density plane.
These solutions however, may be either unstable or be
energetically unfavourable with respect to solutions in
which replica symmetry is broken. To analyse these 
issues in detail requires a choice of potential and a numerical
solution, which we give for certain examples in sections
6 and 7. Meanwhile we discuss the general conclusions that
can be drawn from the simple $q=0$, $M=0$ solutions,
the stability criteria and
issues that arise if there is ferromagnetic order.

Given a solution of \MFTRS\ for $q$ and $M$, the correlators
for the spin density operators can simply be determined 
by \MagCorrelator\ using the variational form:
$\langle \phi_a(r)\phi_b(r') \rangle =G_{ab} $.
The correlator for the $q_{ab}$'s is less immediate
and in Appendix A we give the general RS solution of \Qequation .

%%%%%%%%%%%%%%%%%%%%%%%%%%%%%%%%%%%

\subsec{Region with $q=0$, $M=0$.}

The $M$ equation of \MFTRS\ can always be satisfied by setting $M=0$,
and if a solution of the $q$ equation exists then there is a 
paramagnetic phase. 
A particularly simple situation occurs 
in the region of high temperature and low density
where $q=0$ is a unique solution
for a wide class of potentials and dimensions.

In this case, $g_1$ vanishes and the $\phi_a$ propagator 
is diagonal in replica indices.
It follows from \MagCorrelator\ that the magnetisation correlator 
is also diagonal.
By using the results of appendix A for $q=0$, or by noticing 
that for diagonal $G_{ab}$, the second equation \Qequation\ simplifies
to give $\tilde \Sigma_{abcd} = \Omega_{abcd} \int {d^dp\over (2\pi)^d}\tilde 
g_0(p)\tilde g_0(k-p)$, the $q_{ab}$ correlator can also be found,
\eqn\CorrelatorsRSdiag{\eqalign{
\langle \tilde M_a(k) \tilde M_a(-k)\rangle &=
{\rho\over \left(1-\rho\beta \tilde J(k)\right)}\cr
\langle \tilde q_{ab}(k) \tilde q_{ab}(-k) \rangle &=
{\rho \over \left(1-\rho\int {d^dp\over (2\pi)^d}\tilde 
g_0(p)\tilde g_0(k-p)\right)}\cr
}}
All other $q_{ab}$ correlators vanish except for 
$\langle \tilde q_{aa}(k) \tilde q_{bb}(-k) \rangle = \rho$.
The spin correlators follow simply from these formulae
according to \MagPhysical\ and \QSpin.

Both correlators may, depending on the type of interaction,
diverge somewhere in the temperature density  plane. 
As temperature is lowered, 
the divergence in 
$\bigl[ \langle \tilde S(k)\tilde S(-k) \rangle^2\bigr]_{av}$
will always occur before (or at the same point) as the divergence in
$\bigl[ \langle \tilde S(k)\tilde S(-k) \rangle \bigr]_{av}$.
If we assume that the first divergence occurs at zero momentum, and
can therefore correspond to a continuous spin glass
phase transition, the line on which it occurs is given by the
condition:
\eqn\Divcond{
\rho\int {d^dk\over (2\pi)^d} \tilde g_0(k)\tilde g_0(k) = 
{g_1\over q} = 1
}
Where we have used the relation
$\tilde g_1 = \rho q \tilde g_0^2$ \GsolnRS,
which shows that $g_1$ is linear in $q$.
The spin density correlator generally remains finite on this line.
The above condition is identical to
the condition for the first appearance of a non-zero $q$ solution
in the equations \MFTRS, since by 
expanding the right hand side of the $q$ equation (for $M=0$) we have,
\eqn\Qeqnexp{
q =  g_1 - 2 g_1^2 + \smallfrac{17}{3} g_1^3\dots
}
The coincidence of the two lines is not surprising as it 
only relates to two different ways of identifying the line
of the spin glass transition. 

In Appendix B
we present the usual stability analysis along the lines of
the work by de~Almeida and Thouless \AlTh, and find the
following condition that the replicon eigenvalue remain positive,
\eqn\ATcond{
\rho\, r \int {d^dk\over (2\pi)^d}\ \tilde g_0^2 
= {r\over q} g_1 < 1
}
Where $r$ is given by
\eqn\rdefn{
r=
\left( 1 -2\Omega_{bc} +\Omega_{abcd}\right)=
{1\over \sqrt{2\pi}}\int d\xi e^{-{\xi^2\over 2}}
{\rm cosh}^{-4}(\beta \tilde J(0) M +\xi\sqrt{g_1(0)})
}
The solution discussed above, in which $q$ vanishes,
has $r = 1$, and is only stable above the same spin glass 
transition line, now also identified as the AT line. 
Just below the AT line, the perturbative 
RS solution with $q$ small is always unstable.
Performing the same expansion as in \Qeqnexp\ for small 
$g_1$ with $M=0$ for the $r$ equation above, we find
$r = 1 - 2g_1 + 7 g_1^2 + O(g_1^3)$.
The condition \ATcond\ is therefore always violated
and the AT line signals a non-trivial breaking of replica symmetry.
This indicates that we have a situation similar to that
found in the SK model: namely a high temperature $q=0$
phase separated from a spin glass phase with non trivial RSB.
Usually the coincidence of lines noted here indicates that the RSB 
will take place continuously.
The reader will doubtless note that the above analysis seems remarkably
general both in terms of the type of interaction involved and on the
dimensionality of the space. 
One sees that when presented with the statistics of the matrix 
$J(r_i - r_j)$ it is very difficult to divine from what interaction type and
what dimension of space it arises from. The consideration of a 
genuine cut off interaction where one would expect the geometry 
and type of interaction to play a much stronger role is relegated 
to a future study.

%%%%%%%%%%%%%%%%%%%%%%%%%%%%%%%%%%%%

\subsec{Magnetisation}

For a ferromagnetic interaction the mean field equations, 
$M = \rho\tanh (\beta\tilde J(0)M)$ suggest that we should expect
a ferromagnetic transition at $\rho\beta\tilde J(0) = 1$.
The Gaussian variational approximation sees this
transition as the divergence in the $M_a$ correlator in 
\CorrelatorsRSdiag.
However it is clear from the preceding discussion that 
the spin glass transition intervenes and
that we should not even be attempting to solve the RS equations
\MFTRS\ to locate a ferromagnetic transition because the 
system is already in an RSB phase. The AT line shrouds
the critical region and the RS equations are not
relevant to the transition.

Such a phenomenon may be surprising and the order in which 
the transitions occur deserves further discussion.
For a generic interaction and generic number of spin
components it is reasonable for the spin glass transition to 
occur at higher temperature than the ferromagnetic one. 
For example this is known to happen
in the random field Ising model, where replica symmetry
breaking has been shown to take place
before the ferromagnetic transition \Remi.
The problem occurs for Ising spins and a purely ferromagnetic interaction
(positive in space) where we might well
expect the transitions to be concurrent.

Let us try and clarify this point \ThankG\ by noting that the divergence in 
$\bigl[ \langle \tilde S(k)\tilde S(-k) \rangle^2\bigr]_{av}$
occurs while 
$\bigl[ \langle \tilde S(k)\tilde S(-k) \rangle\bigr]_{av}$
is still finite.
The divergence implies that at sufficiently large spatial separation
the spin glass correlator is larger than the ferromagnetic one:
$\bigl[ \langle S(r) S(0) \rangle^2 \bigr]_{av}
>\bigl[ \langle S(r) S(0) \rangle \bigr]_{av}$
We might expect that these correlators are distributed according
to some probability distribution $P(C)$ and therefore write,
\eqn\Ineqality{\eqalign{
\Bigl[ \langle S(r) S(0) \rangle \Bigr]_{av}
&= \int P(C) C dC\cr
\Bigl[ \langle S(r) S(0) \rangle^2 \Bigr]_{av}
&= \int P(C) C^2 dC\cr}}
But for Ising spins we expect that the correlator
$C = \langle S(r) S(0)\rangle$ is bounded,
$0\le C\le 1$, provided all the interactions are ferromagnetic.
Using such bounds in the expressions above, it is clear that
the inequality,
$\bigl[ \langle S(r) S(0) \rangle^2 \bigr]_{av}
\leq \bigl[ \langle S(r) S(0) \rangle \bigr]_{av}$
holds for all $r$.
This inequality is violated in our solutions.
It should be possible to trace this breakdown of the Gaussian variational
approximation to neglect of some higher order (in number of spin 
components) term in the Schwinger Dyson equations.
Indeed, Sherrington \She\ has pointed out the dangers inherent
in methods similar to the one we use in the case of a 
Landau Ginzburg approach to dilute ferromagnets, and has 
identified the relevant diagrams.

Even though it is inappropriate to look for a continuous
ferromagnetic transition in the RS equations, 
it is interesting to see what happens.
The usual argument, based on the graphical solution
of \MFTRS, provides a simple condition 
on the gradient of the magnetisation equation (at $M= 0$)
for the existence of small $M$ solutions:
$(1-q)\rho\beta \tilde J(0) > 1$
(the $(1-q)$ term causes a suppression of the transition
temperature with respect to the mean field prediction).
On the other hand, for the Gaussian integrals to be well defined 
the propagators must be positive leading to,
$(1-q)\rho\beta \tilde J(k) < 1, \quad \forall k$.
In this sense the usual condition always fails, even in finite volume.
This argument does not preclude RS solutions with magnetisation,
it only indicates that the ferromagnetic transition itself is hidden,
or possibly first order.
Indeed for suitable ferromagnetic interactions 
numerical analysis of the equations allows us to find ferromagnetic
solutions below the transition with no difficulty.
At zero temperature general arguments would require such a solution 
for interactions that do not vanish beyond some range.
Since this issue requires a choice of potential
we will discuss the issue further when we look at examples.

%%%%%%%%%%%%%%%%%%%%%%%%%%%%%%%%%%%%

\newsec{Replica Symmetric Breaking}

Since we have found a scenario similar to that of the
Sherrington Kirkpatrick model, 
in which the RS $q\ne 0$ solution is unstable as soon as it appears,
we shall look for continuous replica symmetry broken solutions.
We consider Parisi matrices
in which the off-diagonal part of the matrix $\tilde G_{ab}(k)$
is parameterised by a continuous function $\tilde g(u,k)$
where $u\in [0,1]$, and the diagonal part is denoted by $\tilde g_D(k)$.
The algebra of such matrices has been developed, and
expressions for products, inverses {\it etc} are given in the
appendix of reference \GandM.
We shall follow the notation used in that paper for certain 
integral transforms that occur frequently.
Although the similarity to the random manifold problem seems
clear the sign of the potential is opposite and 
is related to a random manifold problem with imaginary noise.

For $G_{ab}$ a Parisi matrix, $\Omega$ \Omegadef,
is very similar to the free energy in the SK model.
It is well known that this cannot 
be obtained in a closed form and a standard strategy is to
work close to the transition line by expanding $\Omega$ up to a term of 
$O(g^4)$ which in the SK model is the first term leading to a breaking
of replica symmetry \Pa. 
We shall assume that there is no magnetisation in this calculation because
the denominators of the magnetisation correlator showed no tendency
to diverge at the AT line.
The expansion is, \PytteR,
\eqn\OmegaRSB{ 
{\Omega-1\over n} \approx  {1\over 2} g_D(0) 
- {1\over 4} \int_0^1 \left( 
g^2 + {1\over 6} g^4 -{u\over 3} g^3 - g\int_0^u g^2 \right) \ du
}
where $g$ is the spatial function, $g(u,0)$, evaluated at zero distance.
The remaining terms in the action are easily computed within the algebra
of Parisi matrices. The variational equations are,
\eqn\VareqnsRSB{\eqalign{
[\tilde g_D(k)]^{-1} &= + \rho \sigma_D(k) 
= \rho \left( {\tilde J^{-1}\over \beta\rho} - 1 \right) \cr
[\tilde g(u,k)]^{-1} &= - \rho \sigma (u) 
= 2\rho {\delta \Omega\over \delta\tilde g}\cr}
}
Where the notation $[\tilde g]^{-1}$ indicates the component of
the inverse matrix which is also of Parisi form.
In deriving these equations of motion it is important to take
care of the signs; for example a minus sign appears in
$\delta/\delta\tilde g(u,k) \, {\rm Tr} \log G = - [\tilde g(u,k)]^{-1}$,
and the signs for $\sigma$ have been chosen 
in order that the following equations for the inverses 
appear in the simplest form. For the antiferromagnetic sign interaction
the definition \OmegaRSB\ is changed and a sign appears in the second
of the variational equations.

Defining the denominator combinations:
\eqn\Denominators{\eqalign{
D_D(k) &= \sigma_D + \langle \sigma \rangle , \cr 
D(u,k) &= \sigma_D + \langle \sigma \rangle + [\sigma ](u),\cr}
}
where following the notation of \GandM\ the angle and square brackets 
denote 
$\langle \sigma \rangle \equiv \int_0^1 \! du\sigma (u)$
and
$[\sigma ](u) \equiv u\sigma (u) -\int_0^u \! dv\sigma (v)$.
The equations can be inverted to find,
\eqn\InvertRSB{\eqalign{
\tilde g_D(k) &= {1\over \rho D_D(k)} \left( 1 + \int_0^1\! {du\over u^2}
 {[\sigma ](u) \over D(u,k)} + {\sigma(0)\over  D_D(k)}\right)\cr
\tilde g(u,k) &= {1\over \rho D_D(k)} 
\left( {[\sigma ](u) \over u D(u,k)} + \int_0^u\!{dv\over v^2}
 {[\sigma ](v) \over D(v,k)} + {\sigma(0)\over  D_D(k)}\right).\cr}
}
Taking the derivative with respect to $u$ 
of the second of these equations leads to the simple relation,
\eqn\gderiv{
\tilde g' = {1\over \rho}{\sigma'\over D^2} 
=-{1\over u\rho}{d\over du}{1\over D} \ .
}
where we have made use of $D' = [\sigma]' = u\sigma'$.
Proceeding by differentiating the second equation of \VareqnsRSB\ 
with respect to $u$ and using the expression for $g'$ above,
one obtains $\sigma ' = 0$, or,
\eqn\Derivone{
\left( 1+  g^2 - u g  - \int_u^1 g \right) = 
\rho \left( \int\!  {d^dk\over (2\pi)^d} {1\over D^2} \right)^{-1}\ .
}
Taking another derivative in some region where equation
\Derivone\ holds we find:
\eqn\Derivtwo{
g = \alpha (u) u 
= {u\over 2} \left( 1 + 2\rho^2 
{\int  {d^dk\over (2\pi)^d} {1\over D^3} \over 
\left( \int  {d^dk\over (2\pi)^d} {1\over D^2} \right)^3} \right).
}
\ifig\RSBpattern{
Typical RSB pattern just below the AT line.}
{\epsfysize=6cm \epsfbox{siteRSB.eps}}
Substituting this result into the relation \gderiv\ 
we find that $\sigma$ is determined by a 
a non-linear, but first order, differential equation in $u$,
\eqn\Diffeqn{
{d\over du}\Bigl( u\alpha (u) \Bigr)
= - {1\over u\rho}\int {d^dk\over (2\pi)^d}{d\over du} {1\over D}
}
We cannot go further without being more explicit about the form of the 
interaction and dimension. 
In general the function $\alpha(u)$ 
has a complicated dependence on $\sigma(u)$ and the differential equation
will be difficult to solve, but based on the examples we have
studied, we expect that the RSB pattern as shown in 
\RSBpattern\ to be generic.
The initial part of $g(u,0)$ is determined by
a power series analysis of the nonlinear 
differential equation about the origin, and in all examples we have 
studied the leading term is linear.
Above some breakpoint, $u_0$ close to the origin, $g(u,0)$ becomes
constant. For consistency with the original expansion of 
$\Omega$ we must work close to the AT line and must require that
the $g(u,0)$ is perturbatively small in deviations from that line.
Other patterns could be 
envisaged, though we have not yet found any examples.

For certain choices of interaction or dimension the differential equation
may simplify radically. For example in large enough dimensions 
(the critical dimension depends on the large momentum behaviour
of the interaction; if  $\tilde J \sim k^{-2}$, then $d_{crit}=4$) 
the second term in $\alpha$ will disappear 
in the limit in which the short distance cutoff is removed.
The resulting equation is simple and we find a
scenario very similar to that found in the SK model.
Due to a remarkable cancelation 
$\alpha$ is also a constant for
Yukawa interactions in 3 dimensions. 

This analysis near the AT line is insufficient for some purposes
such as to investigate continuous RSB solutions with magnetisation,
and one would like to solve the equations in greater generality.
More terms could be kept in the expansion \OmegaRSB\ 
\PytteR \ThAlKo, or 
it may be possible to use the differential equation obeyed by $\Omega$
along the lines of the SK case \PaDu.
Further difficulties would occur if it became necessary to
consider the order parameters with more replica indices 
that can be constructed as in section 2.1 \ViBr.
We do not attempt such generalisations here.

We have also studied the equations for one step of replica symmetry 
breaking. We do not expect such solutions to be 
relevant near the AT line, and at lower temperatures 
their importance should be judged relative to some continuous
solution. Another use of such solutions is in calculating the
entropy of the meta-stable states which is relevant to the dynamics.
We have made a calculation following Monasson \Remib\ that demonstrates
that as usual in SK-like situations, this entropy remains zero down to the
AT line and that we therefore do not expect any distinct 
dynamical transition.

%%%%%%%%%%%%%%%%%%%%%%%%%%%%%%%%%%%%%%%%%

\newsec{Yukawa Potential.}

A simple potential for illustrative purposes is, 
\eqn\Yukawa{
\tilde J(k) = \pm{1\over \mu^{d-2}}{1\over k^2 + \mu^2}
}
where the signs refer to ferromagnetic and antiferromagnetic
interactions respectively.
This momentum space form describes a Yukawa type potential that is
screened on a length scale $1/\mu$. We choose to measure distances
in terms of this scale (for example the 
dimensionless density becomes $\rho/\mu^d$),
and therefore set $\mu =1$ in what follows.

The ferromagnetic sign leads to a rich phase diagram
and the complications discussed in section 4.2. For this reason it is 
useful to consider the antiferromagnetic sign 
since antiferromagnetic order cannot exist
in the absence of a lattice, and the phase diagram will be 
simpler, allowing us to concentrate on the properties of the
spin glass transition.

We discuss different dimensions separately, starting in three dimensions.

%%%%%%%%%%

\subsec{Three dimensions}

In three dimensions the real space form of \Yukawa\ is the well 
known Yukawa form,
\eqn\YukawathreeD{
J(r) = \pm{e^{-\mu r}\over 4\pi \mu r}
}
In this case the integral \GoneRS\ can be done to obtain $g_1$:
\eqn\GthreeD{
g_1 = {\rho \beta^2 q}\int {d^3k\over (2\pi)^3} 
{1\over (k^2 + 1 \mp(1-q)\rho\beta)^2}
= {\rho \beta^2 q\over 8\pi \sqrt{1 \mp(1-q)\rho\beta}}
}
which allows us to numerically solve the equations \MFTRS.
At high temperature the unique solution is $q=0,M =0$,
whereas at lower temperatures for the ferromagnet, the positivity of $g_0$
restricts the possible range of $q$ to,
$q> 1-1/\rho\beta$.

The spin glass transition line is defined by the condition
$8\pi\sqrt{1\mp\beta\rho} = \rho\beta^2$, which can be solved to give,
\eqn\YukATline{
\rho = 2(4\pi T)^2(\mp T+\sqrt{T^2+1/(4\pi)^2}) \ .
}
This is the only phase transition in the antiferromagnetic case
whereas in the ferromagnet it lies very close to the line, $\rho\beta = 1$,
where ferromagnetic ordering would occur in mean field approximation.

Above this line, in the high temperature phase with $q=0,M =0$,
only the diagonal magnetisation correlator remains:
\eqn\YukawaMagcorr{
\langle M_a(r)M_a(0)\rangle
 = \rho \delta(r) + {\rho^2\beta e^{-r\sqrt{1-\rho\beta}}
\over 4\pi r}
}
We have given the real space form because it has a simple interpretation.
The first term is from the same spin, 
and the leading piece of the second term arises from a two spin term.
Only considering two spins, the correlator can be written in terms of
Boltzmann factors which when expanded for small interaction strength
give $\rho^2\beta J(r)$. Using \YukawathreeD\ one recovers
the leading piece of the full correlator.

Still in the high temperature region,
we can also calculate the only non-trivial $q_{ab}$  correlator for the
$P$ combination of indices (see Appendix A):
\eqn\YukawaQcorr{
\tilde Q_{abab}(k)
=\langle \tilde q_{ab}(k) \tilde q_{ab}(-k) \rangle
={\rho\over 1 - \rho\int{d^dk\over (2\pi)^3} \tilde g_0^2(k)}
={\rho\over 1 - {\rho\beta^2\over 4\pi} {1\over k} \sin^{-1}
\left({k\over \sqrt{4(1\mp \beta\rho)+2k^2}}\right)}
}
The connected spin correlation function 
$\bigl[ \langle \tilde S(k)\tilde S(-k) \rangle^2_{con} \bigr]_{av}$
is obtained by subtracting the term 
$\langle \tilde q_{aa}(k) \tilde q_{bb}(-k) \rangle = \rho$ \QSpin.
In the small momentum limit this expression simplifies and becomes
proportional to $1/(k^2 + m^2)$ with mass given by,
\eqn\YukawaQmass{
m^2= {96\pi\over 5\rho\beta^2}(1-\beta\rho)
\left(  \sqrt{1-\beta\rho} - {\rho\beta^2\over 8\pi}\right)
}
In accordance with the general analysis, criticality occurs
for vanishing $m^2$ which is another way of finding the AT line.
Despite the fact that we are employing a Hartree Fock method,
which in pure models can give non-classical exponents,
here $m^2$ is an analytic expression and the exponent $\nu$
takes its mean-field value of $\smallfrac{1}{2}$. 
\ifig\FthreeDdiag{
The regions in the density temperature plane where various
RS solutions exist. 
For ferromagnetic Yukawa interactions in 3 dimensions. A region
between the spin glass and ferromagnetic transitions is unphysical.}
{\epsfysize=6cm \epsfbox{F3Dlines.eps}}
At lower temperatures, other replica symmetric solutions exist. 
The simplest case is the
antiferromagnet for which there are two solutions below the AT
line. One solution, $q=0$, is unstable both with respect to
longitudinal and replicon fluctuations, and the other has $q > 0$
and is numerically found to be unstable with respect to replicon
fluctuations as expected from the general analysis in section 4.2. 
The ferromagnetic case is more complicated, besides
$M=0$ solutions (the $q=0$ solution only exists for $\beta\rho \leq 1$)
that are unstable just as for the 
antiferromagnet,
there are solutions with $M\ne 0$.
These magnetised solutions first appear on a second line at slightly
lower temperature than
the AT line, as shown in \FthreeDdiag . On this line there is one new
stable solution with non-vanishing magnetisation and $q$,
these values being largest at the low temperature end of the curve. 
As the temperature is reduced the solution splits into two; 
the branch with larger values of $M$ and $q$ remains stable and the 
magnetisation continues to grow in the usual way, the other
branch decreases in magnetisation and becomes unstable with
respect to the replicon not far below the line shown on the figure.
We suspect that the decreasing solution will not play any role.
%The vigilant reader gifted with extra-sensory perception 

The significance of the upper line in \FthreeDdiag\
and the new solutions that it heralds
depends on whether continuous solutions also exist in this region. 
Free energies must then be compared to identify a first order transition. 
As was discussed in section 4.2, for purely ferromagnetic interactions 
we expect concurrent spin glass and ferromagnetic
transition lines, and therefore we are witnessing a shortcoming
of the approximation.
It seems likely that at sufficiently low
temperature the single stable new RS ferromagnetic solution will
have the lowest free energy and will represent the system.
If this is the case, at zero temperature the transition must
occur for finite density, suggesting a connection between the 
ferromagnetic and a percolation transition. 
This would be a reasonable conclusion for a
ferromagnetic potential that strictly vanishes beyond some radius
(it would be interesting to consider such a potential
to clarify this connection).
On the other hand, for an interaction of the type \YukawathreeD\
which never vanishes, we should always 
expect a ferromagnetic RS ground state at zero temperature.
Investigating this region in detail, we find that
the upper line shown in \FthreeDdiag\ does not quite extend 
down to zero temperature because the main magnetic solution becomes 
unstable with respect to replica fluctuations.
This observation can be confirmed by making a low temperature 
expansion of the RS equations \MFTRS. Apparently this is another 
failure of the approximation, and is maybe not surprising
since it is unlikely that the $\beta \rightarrow \infty$
limit commutes with the large number of components limit
used to justify the approximation.

\bigskip

We now turn to the form of continuous RSB just below the spin glass 
transition.
In view of the arguments showing that the critical region for
the pure ferromagnet is not treated well by the approximation,
we only discuss the antiferromagnet.

It is useful to identify mass parameters by writing the denominators
\Denominators\ as,
\eqn\YukawaRSBdenom{\eqalign{
D_D(k) &= {1\over \beta \rho} \left( k^2 + m_D^2 \right)\cr
m_D^2 &= \Bigl( 1 + \beta\rho 
\bigl(1 + \langle \sigma\rangle\bigr)\Bigr)\cr
D(u,k) &= {1\over \beta \rho} \left( k^2 + m^2 (u) \right)\cr
m^2 (u) &= m_D^2 + \beta\rho[\sigma ](u)\cr}
}
Performing the momentum integrals over inverse powers of the denominator we
find a remarkable cancelation so that the parameter $\alpha$ \Derivtwo\ is
independent of $u$,
\eqn\YukawaRSBalpha{
\alpha = {1\over 2}\left({32\pi^2 \over \rho \beta^3} - 1\right)
}
We consider solutions consisting of a continuous piece 
up to a breakpoint $u_0$, followed by
a constant as shown in \RSBpattern.
The differential equation is trivially solved to give
$m(u) = m_D + 2\pi\alpha u^2/\beta$. Continuing to solve the
RSB equations we find that in the continuous
region, $\sigma$ contains linear and cubic terms in $u$
and that the breakpoint is determined by a quartic equation.
Since $u_0$ is small we only give the leading terms,
\eqn\YukawaRSBquartic{
{\alpha\over \beta} (32\pi^2 \pm \rho\beta^3)u_0
=\left(\rho\beta^2 -8\pi\sqrt{1 + \rho\beta}\right)\ .
}
The right hand side vanishes on the AT line, recovering the RS solution
with $u_0 = 0$.
Expanding to first order about this line we find an expression for the
breakpoint in perturbation theory,
\eqn\YukawaRSBbreak{
 u_0 = 4{\delta \beta\over \beta} 
{(\rho\beta^3) (\rho\beta^3 - 16\pi^2)\over (\rho\beta^3 - 32\pi^2)^2}
}
This indicates that we have a consistent solution since
$g$ remains perturbatively small so the expansion near the AT line is
justified.
The magnetisation correlator can be calculated in this region
along the lines of \MagPhysical, in which the sum now becomes an integral
\InvertRSB.
To leading order we find
\eqn\RSBgdiag{
\tilde g_D(k) = {\beta\over k^2 + m_D^2}
\left( 1 + {4\pi\alpha m_D\over \beta (k^2 + m_D^2)}\right)
}
Where $m_D = \sqrt{1 + \rho\beta} + O(u_0)$. 

These detailed predictions of a spin glass phase for a site
disordered antiferromagnet are interesting because few analytic
results exist for the problem. Preliminary numerical simulations
of the dynamics \Giulia\ reach the same conclusion as those
described by McLenaghan and Sherrington \McSh\ (who also refer
to experimental realisations); that is to say, no such phase is 
observed.  More work is needed in this area since for the analytic work,
either the Gaussian variational approximation or the simple choice 
of a two replica order parameter may be in doubt; whereas the
numerical simulations suffer from small sizes and consequent 
dependence on the boundary conditions.

%%%%%%%%%%%%%%%%%%%%%%%%%%%%%%%%%%%%%

\subsec{Yukawa in  $d > 4$.} 

In four or more dimensions many of the integrals diverge indicating
dependence on details of an unknown short-distance theory.
For example, in the RS phase the integral defining $g_1(0)$ diverges, the 
leading term (in the momentum cutoff $\Lambda$) gives:
\eqn\YukawafourG{
g_1 = \rho \beta^2 q {\Gamma_d\over (d-4)} 
\left( {\Lambda\over \mu}\right)^{d-4}
}
Where $\Gamma_d= {\Omega_d/(2\pi)^d = 2/\bigl((4\pi)^{d/2}\Gamma (d/2)\bigr)}$.
The AT line is thus given by,
\eqn\YukawafourAT{
\rho \beta^2 = {(d-4)\over \Gamma_d}
\left( {\mu\over \Lambda}\right)^{d-4}
}
So as the cutoff is removed ($\Lambda \to \infty$), 
the AT line moves towards the $\rho = 0$ axis,
and the RSB phase covers the whole phase space.
One can proceed to expand about this line
to find a simple RSB with $\alpha = 1/2$
and $u_0 =  4\delta\beta /\beta$.

Alternatively, we can recover a finite temperature transition  
by taking the infinite dimension limit
while scaling the cutoff as  $\Lambda/\mu \sim \sqrt{d}$.
This limit sets $g_1 \propto \beta^2 q J^2$ and yields the SK mean field
equations.

%%%%%%%%%%%%%%%%%%%%%%%%%%%%%%%%%%%%%

\subsec{Yukawa in  $d < 3$.} 

The two dimensional example contains most of the features seen in low
dimensional Yukawa models, and we concentrate on it here.
Again the integral can be done to obtain a simple form for $g_1$,
\eqn\GtwoD{
g_1 = {\rho \beta^2 q}\int {d^2k\over (2\pi)^2} 
{1\over (k^2 + 1 \mp(1-q)\rho\beta)^2}
= {\rho \beta^2 q\over 4\pi \left( 1 \mp(1-q)\rho\beta\right)}
}
The AT condition is 
$4\pi(1\mp \rho \beta) =\rho \beta^2$, so the line is given by:
\eqn\YuklowDAT{
\rho = {4\pi T^2\over 1 \pm 4\pi T}
}
Note that in the antiferromagnetic case, the line always 
lies below $T= 1/4\pi$.
Numerical investigations of the RS equations lead to a picture
qualitatively the same as the one described in the three dimensional case.
Quantitatively there are differences, notably,
in two dimensions the unphysical region between the AT line and the line
on which stable ferromagnetic solutions first appear,
is wider than in \FthreeDdiag.

In searching for a RSB solution for the antiferromagnetic case 
it is useful to work in terms of the mass parameter $m(u)$,
defined by the combinations \YukawaRSBdenom\ rather than directly
in terms of $\sigma$.
In two dimensions we find
$\alpha = {1\over 2}({16\pi^2\over \rho \beta^3}m^2 - 1)$.
The differential equation \Diffeqn\ can be written in terms of the 
variable $x = u^2$, and it is possible to work in terms of 
$f = m^2$, to obtain,
\eqn\YuktwoDE{
\left( {32\pi^2\over \rho \beta^3}x f - {\beta \over\pi}\right)
{df\over dx}
+
\left({16\pi^2\over \rho \beta^3}f - 1\right)f
= 0
}
Provided the breaking pattern is similar to that shown in \RSBpattern\
we need only analyse the small $u$ behaviour of this equation.
We find results very similar to the three dimensional case:
$m^2 \approx m_D^2(1 + \pi u^2(16\pi^2m_D^2 - \rho\beta^3)/\rho\beta^4)$,
and $g$ approximately linear in $u$.
This is a consistent procedure since near the AT line we 
obtain a sensible expression for the breakpoint,
\eqn\Yuktwobreak{
u_0 = 4\delta \beta 
{(\beta - 2\pi)\over (\beta - 4\pi)^2}\ .
}

Closer inspection of the differential equation shows that it
is ill-behaved because
solutions tend to blow up at finite values of the parameter $u$.
Also note that two distinct, constant solutions exist,
which may suggest a solution with one step of RSB.

One dimension is interesting because
exact results are known about the bond disordered version of the model.
The AT line is given by $4(1\mp\rho\beta)^{3/2} = \rho\beta^2$,
which is a cubic equation in $\rho$.
A series solution of the differential equation
shows behaviour like that of the two dimensional case.
The zero dimensional case is formally similar to an RKKY
potential in any dimension, and is better discussed in that context.

%%%%%%%%%%%%%%%%%%%%%%%%%%%%%%%%%%%%%%%%%

\newsec{RKKY-like Potential}

The RKKY potential itself, which in three dimensions
is $J(r) = {\cos \mu r/ r^3}$,
has a complicated Fourier transform. 
So as is usual in analytic work we consider instead:
\eqn\RKKYJ{
\tilde J(k) = \cases{\tilde J_0 /\mu^d, &for $k<\mu$;\cr
                     0, &for $k>\mu$.\cr}
}
$J(r)$ oscillates in sign to have the correct qualitative form,
but note that the negative parts are not very strong.
As was the case with the Yukawa potential, 
$\mu$ and $\tilde J_0$ 
can be scaled away by measuring distances and temperatures appropriately,
so we will set $\mu= \tilde J_0 = 1$.
The integral for $g_1$ can immediately be performed, 
\eqn\RKKYG{
g_1 = 
{\rho \beta^2q}\int_{|k|<\mu} {d^dk\over (2\pi)^d} 
{1\over \left(1-(1-q)\rho\beta \right)^2}
=
{\Gamma_d\rho \beta^2 q\over \left(1-(1-q)\rho\beta \right)^2}
}
Where $\Gamma_d= {\Omega_d/(2\pi)^d = 2/((4\pi)^{d/2}\Gamma (d/2))}$,
and in three dimensions, $\Gamma_3 = 1/6\pi^2$.
\ifig\RKthreelines{
Regions of existence of RS solutions 
for RKKY--like interactions in three dimensions.}
{\epsfysize=6cm \epsfbox{RK3Dlines.eps}}
The AT condition is given by,
\eqn\RKKYATcond{
{\Gamma_d \rho \beta^2\over \left(1-(1-q)\rho\beta \right)^2}=1
}
So at this transition the magnetisation propagators remain finite.
This can be solved to give the AT line,
\eqn\RKKYATline{
\rho = (T + {\Gamma_d\over 2} - \sqrt{\Gamma_d T + {\Gamma_d^2\over 4}})
}

Since the form of this potential does not depend strongly on the
dimension we shall discuss only the three dimensional case.
Above the AT line, in the $q=0$ RS phase, the correlators are
\eqn\RKKYMagcorr{
\langle \tilde M_a(k)\tilde M_a(-k)\rangle
=\cases
{{\rho\over 1 - \rho\beta} & for $k< 1$;\cr
\rho & for $k> 1$;\cr}
}
and
\eqn\RKKYQcorr{
\tilde Q_{(ab)(ab)}(k)
=\cases
{{\rho\over 1 - {\rho\over 96\pi^2} \left({\beta\over1-\rho\beta}\right)^2
(k+4)(k-2)^2},
 & for \hbox{$k < 2$};\cr
\rho, & for \hbox{$ k> 2$}.\cr}
}
Where the non-trivial form of the second propagator arises from
the slightly complicated integration region and this allows us to obtain 
a non-mean field exponent, $\eta = 1$.

Numerical study of the RS equations exposes a scenario very similar to the
ferromagnetic Yukawa case. The solutions with zero magnetisation
are always unstable below the AT line, but new magnetised solutions
appear on another line. 
We recognise the second line from the ferromagnetic experience,
at very high densities the positive short range part of the potential
dominates and the system magnetises, returning to a RS form.
The phase diagram is shown in \RKthreelines.

For RSB we find
$\alpha = {1\over 2}(1 + {2\over \Gamma_d^2 \rho \beta^3}m^6)$.
The differential equation can be written in terms of $f= m^2$ where 
$m$ is defined in a similar way to \YukawaRSBdenom ,
\eqn\RKKYDE{
\left( {12\over \Gamma_d^2\rho \beta^3}x f^4 - {4\beta\Gamma_d}\right)
{df\over dx}
+
\left({2\over \Gamma_d^2\rho \beta^3}f^3 + 1\right)f^2
= 0
}
The expression for the breakpoint is sensible and given by,
\eqn\RKthreebreak{
u_0 = 4 \Gamma_d\delta \beta 
{\rho \beta\over (1 + \rho\beta)(1 - \rho\beta)}\ .
}
From this information we can calculate the structure of the two
index correlators $\tilde g(k,u)$, leading to non trivial 
momentum dependence in $\tilde g_D(k)$, however 
the analysis only holds close to the spin glass transition 
and cannot address the ferromagnetic transition itself.
Indeed, at the ferromagnetic line, $g$ will be large and
the expansion of $\Omega$ is manifestly inappropriate.
In addition the four index correlation functions $Q_{abcd}$
contain much of the physics of the spin glass phase:
for example the $\theta$-exponent \FiHu\
may be extracted from the long distance behaviour of such objects. 
Equation \Qequation\ is however an equation carrying 
four replica indices and the solution in the case of continuous 
replica symmetry breaking is technically rather formidable
requiring extensions of the methods described in \Kondor.

%%%%%%%%%%%%%%%%%%%%%%%%%%%%%%%%%%%%%%%%%

\newsec{Conclusions}

The Gaussian variational method, which is the simplest 
non-trivial way of analysing this site disordered field theory,
provides an interesting picture of the model.
For all forms of interaction between the spins
we find a spin glass transition which separates a 
high temperature paramagnetic phase from a low temperature
phase with non-vanishing Edwards Anderson order parameter. 
The transition is signaled by replica symmetry breaking,
which takes place continuously just below the transition.
We have carefully displayed the connection between physical
observables and the replicated quantities we calculate and
see explicitly how the spin glass correlator,
$\langle \tilde q_{ab}(k) \tilde q_{cd}(-k) \rangle$,
diverges as the transition is approached from above.

The properties of the low temperature region 
depend on the precise form of the interaction,
but not sensitively on dimension in fewer than four dimensions.
Our study of antiferromagnetic Yukawa interactions indicates
that the low temperature phase  
breaks replica symmetry continuously at all temperatures.
The detailed form of the breaking has, however, only been 
determined in the vicinity of the transition.
This prediction is interesting in view of the numerical
simulations which contradict it and further work is
needed in this case.
In the case of ferromagnetic interactions, or the RKKY--like
interactions which contain a substantial ferromagnetic component,
we have found stable replica symmetric solutions with non-vanishing
magnetisation. It seems likely that at sufficiently low temperature
one such solution will describe the system. We are unable to 
be certain on this point until we know about the existence and
properties of solutions with continuous replica symmetry breaking
in the region away from the spin glass transition line.
This problem also means that we are not presently able to
directly analyse the ferromagnetic transition, though,
because the magnetisation correlators remain finite at
the spin glass transition, it is clear that the transitions are
not coincident indicating a breakdown of the approximation
in the case of a pure ferromagnet.
Some further difficulties with the zero temperature limit
have been mentioned in the text.

Besides the inaccessibility of the ferromagnetic transition,
an important aspect of the spin glass physics is missing.
Namely, the behaviour of the spin glass correlator below the 
transition where it should remain massless yet have a 
non-trivial exponent $\theta$ \FiHu. We have derived equations
obeyed by the correlator, but because it is a four index quantity,
the technical difficulties of solving the equation in the 
continuously broken spin glass phase are presently beyond us.
Both the problems mentioned can be addressed in the case
of $m$-component Heisenberg spins in the limit $m\to \infty$ \largeN.

The two issues in which the approximation is clearly unreliable
are in its treatment of the purely ferromagnetic interaction and
it relative insensitivity to dimension or interaction type.
It therefore seems appropriate to add some further words on when
we expect the approximation to be trustworthy.
The Gaussian variational approximation
can be seen as making a Gaussian ansatz on the Schwinger--Dyson
equations of the theory for the fields $\phi_a(x)$ (in the case where there
is RSB the anzatz is actually more refined in terms of nested
Gaussians). However this ansatz can only be justified in certain circumstances.
Some form of the central limit theorem must be brought into play, and the
possible ways in which it would be justified are the following:
\item{$\bullet$} There are many effective neighbours for each spin giving a 
large sum
contributing to the local field. This will be true in a number of cases
such as large spatial dimension (where one is of course closer to mean 
field) and long range interactions for which even in dilute systems
there are many effective neighbours. Therefore one would expect that for 
certain short range applications the theory may be expected to fail.
\item{$\bullet$} When there are long correlations on the local field 
distribution 
one expects the central limit theorem and thus the approximation
not to be applicable. 
Indeed for purely ferromagnetic interactions the correlation between local
fields will become stronger as the temperature is reduced, we see the mass 
in the two-point correlator does become rather small in three dimensions
even if it does not become zero before the predicted spin glass transition.
In the non-ferromagnetic cases the correlations in the local fields do
not become so strong and hence one may have more confidence in the method.

It is natural to enquire what we could say about the
system beyond the level of the Gaussian variational analysis.
Simulations of the basic Hamiltonian we started with 
are at an early stage \Giulia, and the non-trivial dynamics
usually associated with disordered systems has not yet been observed.
The renormalisation group could yield important information about 
the model. A naive application addresses
the ferromagnetic transition because these are the obvious variables
in the problem, but a perturbative approach leads to exactly
the problem investigated by Dotsenko, Harris, Sherrington and Stinchcombe
\DHSS. These authors found that
the renormalisation flow in a dilute ferromagnet was disrupted into
a replica symmetry broken form. The precise interpretation of this
observation is as yet unclear, but we hope that the analysis
we present in this paper based on the Gaussian variational method will shed
light on the problem. One might hope that a more sophisticated approach to
the renormalisation, in which the appropriate variables for the
spin glass transition were isolated, would give new and interesting
information. These issues we postpone to another publication \DeLaRG.

We have observed continuous RSB in finite dimensional models,
and providing an interpretation by identifying the pure states 
in physical terms would be an interesting task. 
For this purpose, 
as mentioned earlier it will be interesting to
consider a potential that strictly vanishes beyond some radius,
as the connection with percolation is clearer and
one would expect to see a greater sensitivity on spatial dimension.

%%%%%%%%%%%%%%%%%%%%%%%%%%%%%%%%

\bigbreak\bigskip\bigskip\centerline{{\bf Acknowledgements}}\nobreak
We would like to acknowledge useful discussions with J.P.~Bouchaud,
M.~Ferrero, G.~Iori, J.~Ruiz-Lorenzo, M.~M\'ezard,  R.~Monasson, 
T.~Nieuwenhuizen, G.~Parisi and P.~Young.

\appendix{A}{The RS correlator for $q_{ab}$}

In this appendix we resolve the equations \Qequation :
\eqn\AQequation{\eqalign{
\tilde Q_{abcd}(k)&= \rho \Omega_{abcd}
+{\rho\over 2}\sum_{gh} \tilde \Sigma_{abgh}(k) \tilde Q_{ghcd}(k)\cr
\tilde\Sigma_{abgh}(k)&= \sum_{ef}\Omega_{abef}\int {d^dp\over (2\pi)^d}
\tilde G_{eg}(p)\tilde G_{fh}(k-p)\cr}
}
for the correlator $\tilde Q_{abcd}(k) = 
\langle \tilde q_{ab}(k) \tilde q_{cd}(-k) \rangle$ in the
general paramagnetic RS case with $M=0$, but arbitrary $q$.

Notice that although symmetry in $a\leftrightarrow b$ 
and $c\leftrightarrow d$ is manifest, the symmetry between pairs 
$(ab)\leftrightarrow (cd)$ is not. This could be rectified by
an appropriate matrix multiplication, but it is simpler to leave.

These equations are a set of linear equations for 
the nine possible index combinations we have
to consider since the diagonal terms are inevitably mixed into 
the equation (the missing symmetry would reduce this to the
seven combinations, $A,B,C,D,P,Q,R$ in the notation of \AlTh).
The four spin trace  $\Omega_{abcd}$ is completely symmetric and
takes values either 1 or $q$ except for the case with all indices different
where it is equal to $-1 +2q +r$. $\Sigma_{abgh}(k)$ 
is defined by quadratics in $g_0$ and $g_1$.
It is definitely not pair symmetric, but does satisfy
$\Sigma_{(aa)(aa)}=\Sigma_{(aa)(bb)}$ and
$\Sigma_{(aa)(ab)}=\Sigma_{(aa)(bc)}$. Other simple combinations 
are based on the longitudinal, $(P-4Q+3R)$, and replicon, $(P-2Q+R)$,
mixtures of 
$P$: $\Sigma_{(ab)(ab)}$,
$Q$: $\Sigma_{(ab)(ac)}$ and
$R$: $\Sigma_{(ab)(bc)}$.

The $9 \times 9$ matrix equation 
splits into a $4\times 4$ and a $5\times 5$ block. 
The $4\times 4$ part is simple giving, 
$\tilde Q_{(aa)(aa)} = \tilde Q_{(aa)(bb)} = \rho$.
The $5\times 5$ part contains the $P$, $Q$ and $R$ pieces and
the longitudinal and
replicon combinations of the Q's are particularly simple,
\eqn\ALongRepl{\eqalign{
\tilde Q_{lon} &= {\rho (-2+2q+3r)\over \tilde D_{lon}} ,
\quad \tilde D_{lon}(k) = 1 - \rho (-2+2q+3r)
\int {d^dp\over (2\pi)^d}\tilde g_0(k-p)(\tilde g_0 - 2\tilde g_1)(p)\cr
\tilde Q_{rep} &= {\rho r\over \tilde D_{rep}} ,
\quad \tilde D_{rep}(k) = 1 - \rho \, r 
\int {d^dp\over (2\pi)^d}\tilde g_0(k-p)\tilde g_0(p)\cr}
}
For the remaining index combinations we find that
the pair symmetry is recovered and that
$\tilde Q_{(aa)(ab)}(k) = \tilde Q_{(ab)(aa)}(k)
=\tilde Q_{(aa)(bc)}(k) = \tilde Q_{(bc)(aa)}(k)$.
The non trivial momentum dependence of these correlators
can be understood since the $q_{aa}$ operator just checks
for the presence of a spin, whereas the  $q_{ab}$ is the squared
magnetisation and obviously depends on the presence of a nearby spin.

The full solution is,
\eqn\AQsoln{\eqalign{
\tilde Q_A & =\tilde Q_{(aa)(aa)} =\tilde Q_B =\tilde Q_{(aa)(bb)} = \rho\cr
\tilde Q_C & =\tilde Q_{(aa)(ab)} =\tilde Q_D =\tilde Q_{(aa)(bc)}
= \rho q + 2\tilde\Sigma_C \tilde Q_{lon}\cr
\tilde Q_P & = \tilde Q_{(ab)(ab)} 
= {1\over \tilde D_{lon}}\left(\rho +
2(\tilde\Sigma_{lon}-\tilde\Sigma_{rep})(\tilde Q_{lon}-3\tilde Q_{rep})
+2\tilde\Sigma_{CD}\tilde Q_C
+2\tilde\Sigma_{R}\tilde Q_{lon}\right)\cr
\tilde Q_Q & = \tilde Q_{(ab)(ac)} 
= {1\over \tilde D_{lon}}\left(\rho +
(\tilde\Sigma_{lon}-2\tilde\Sigma_{rep} + \smallfrac{1}{2})(\tilde Q_{lon}-3\tilde Q_{rep})
+2\tilde\Sigma_{CD}\tilde Q_C
+2\tilde\Sigma_{R}\tilde Q_{lon}\right)\cr
\tilde Q_{R} & = \tilde Q_{(ab)(ac)} 
= \tilde Q_{rep} + {1\over \tilde D_{lon}}\left(\rho +
(-2\tilde\Sigma_{rep} + 1)(\tilde Q_{lon}-3\tilde Q_{rep})
+2\tilde\Sigma_{CD}\tilde Q_C
+2\tilde\Sigma_{R}\tilde Q_{lon}\right)\cr}}
Where besides the longitudinal and replicon components
some of the other $\Sigma$ combinations are needed,
\eqn\ASigma{\eqalign{
\tilde\Sigma_C(k) &=
\rho\int {d^dp\over (2\pi)^d}\tilde g_0(k-p)\tilde g_1(p)
+ \smallfrac{1}{2}
\rho q\int {d^dp\over (2\pi)^d}\tilde g_0(k-p)(\tilde g_0-2\tilde g_1)(p)\cr
\tilde\Sigma_{rep}(k) &=
\smallfrac{1}{2}(1-\tilde D_{rep})\cr
\tilde\Sigma_{lon}(k) &=
\smallfrac{1}{2}(1-\tilde D_{lon})\cr
\tilde\Sigma_{CD}(k) &=
\rho(-2+2q+3r)\int {d^dp\over (2\pi)^d}\tilde g_0(k-p)\tilde g_1(p)\cr
\tilde\Sigma_{R}(k) &=
\rho(2+2q+3r)\int {d^dp\over (2\pi)^d}\tilde g_0(k-p)\tilde g_0(p)
+3\rho(1-q-r)\int {d^dp\over (2\pi)^d}\tilde g_0(k-p)\tilde g_1(p)\cr
&\quad +\smallfrac{1}{2}\rho(-1+2q+r)\int {d^dp\over (2\pi)^d}
\tilde g_1(k-p)\tilde g_1(p)\cr}}
These final forms for each individual component 
are not especially illuminating, but it is worth noting that
they have denominators only of the form $D_{lon}$, $D_{rep}$, $D_{lon}^2$ 
and $D_{rep}D_{lon}$. 

The replicon denominator vanishes first,
\eqn\AATline{
\tilde D_{rep}(k=0) = 1 - \rho \, r\int {d^dp\over (2\pi)^d} \tilde g_0^2(p)
= 1-{r g_1\over q}
}
So we recover the AT condition as the divergence of this correlator.

The case in which $q$ vanishes is particularly simple, 
$\tilde D_{rep} = \tilde D_{lon} = \tilde D$
and only the $P$ combination (and $\tilde Q_{(aa)(bb)} = \rho$)
remains non-vanishing,
\eqn\APcomb{
\tilde Q_{(ab)(ab)}(k)
= {\rho \over \tilde D}
}

Similar techniques could be used to analyse the correlator in
the RS but ferromagnetic phase and also for solutions with one
step of RSB.

%%%%%%%%%%%%%%%%%%%%%%%%%%%%%%%%%%%%

\appendix{B}{Stability of RS solutions}

We present the usual stability analysis based on the eigenvalues for
small fluctuations about a RS solution along the lines of the
work of de~Alemeda and Thouless \AlTh.
We therefore need the Hessian matrix $H_{(ab)(cd)}$:
\eqn\Hessian{
{\delta^2 F_{var}\over \delta G_{ab}(k) \delta G_{cd}(k') }
\propto \left(G_{ac}^{-1}(k)G_{db}^{-1}(k)
+G_{ad}^{-1}(k)G_{bc}^{-1}(k)\right)\delta(k-k')
-{\rho\over (2\pi)^d} \Omega_{abcd}
}
For RS solutions 
the components (in the notation of \AlTh\ 
where all indices are different) of the 
Hessian matrix are given by:
\eqn\HessianPQR{
\eqalign{
A &= H_{(aa)(aa)} 
= 2\left({1\over \tilde g_0} -{\tilde g_1\over \tilde g_0^2}
	\right)^2\delta(k-k')
-{\rho\over (2\pi)^d} \Omega \cr
B &= H_{(aa)(bb)} 
= 2\left({\tilde g_1\over \tilde g_0^2}
	\right)^2\delta(k-k')
-{\rho\over (2\pi)^d} \Omega \cr
C &= H_{(aa)(ab)} = H_{(ab)(aa)} 
= -2\left({\tilde g_1\over \tilde g_0^2}\right)
\left({1\over \tilde g_0} -{\tilde g_1\over \tilde g_0^2}\right)
\delta(k-k')
-{\rho\over (2\pi)^d} \Omega_{ab} \cr
D &= H_{(aa)(bc)} = H_{(bc)(aa)} 
= 2\left({\tilde g_1\over \tilde g_0^2}
	\right)^2\delta(k-k')
-{\rho\over (2\pi)^d} \Omega_{ab} \cr
P &= H_{(ab)(ab)} 
= \left(\left({\tilde g_1\over \tilde g_0^2}\right)^2
 +\left({1\over \tilde g_0} -
{\tilde g_1\over \tilde g_0^2}\right)^2\right)\delta(k-k')
-{\rho\over (2\pi)^d} \Omega \cr
Q &= H_{(ab)(ac)} 
= -\left({\tilde g_1\over \tilde g_0^2}\right)
\left({1\over \tilde g_0} -{2\tilde g_1\over \tilde g_0^2}\right)
\delta(k-k')
-{\rho\over (2\pi)^d} \Omega_{bc} \cr
R &= H_{(ab)(cd)} 
= 2\left({\tilde g_1\over \tilde g_0^2}\right)^2
\delta(k-k')
-{\rho\over (2\pi)^d} \Omega_{abcd} \cr
}}
Diagonalisation in replica space gives the replicon operator
as the combination  $P-2Q+R$.
It is convenient to introduce the positive weight function  $\tilde g_0^{-2}$
into the $k$-space eigen-equation to obtain,
\eqn\Replicon{
{1\over\tilde g_0^2(k)} f(k)
-{\rho\, r\over (2\pi)^d}\int d^dk' f(k') = 
\lambda_{rep}{1\over\tilde g_0^2(k)} f(k)
}
Where $r=\left( 1 -2\Omega_{ab} +\Omega_{abcd}\right)$
is defined in \rdefn.
In the case where $\int f = 0$, the eigenvalue is clearly positive,
otherwise we integrate over $k$ to obtain the condition,
\eqn\ATcond{
\rho\, r \int {d^dk\over (2\pi)^d}\ \tilde g_0^2 
= {r\over q} g_1 < 1
}

The longitudinal mode is slightly more complicated because the
diagonal terms $A,B,C,D$ contribute. The equations for the eigenvalues are,
\eqn\LongABC{\eqalign{
(A-B)f_0 + \Bigl[ (A-B) - 2(C-D) \Bigr]f_1 
&= \lambda_{lon}(f_0+f_1)\cr
2(C-D)f_0 + 2\Bigl[ (C-D) + (P-4Q+3R) \Bigr]f_1 
&= \lambda_{lon}f_1\cr}}
With some rearrangement, and the same weight function, we find,
\eqn\Longitudinal{\eqalign{
{(\tilde g_0 - 2\tilde g_1)\over \tilde g_0}f_0 
+f_1 
&= \lambda_{lon}(f_0+f_1)\cr
f_0 
+\rho (-2 + 2q + 3r) \tilde g_0^2 \int f_1 (k')
&= \lambda_{lon}f_0\cr}}
An inspection of these equations in the same way as for the
replicon leads to the following condition for stability,
\eqn\Longcond{
\rho\, (-2 + 2q + 3r) \int {d^dk\over (2\pi)^d}\ \tilde g_0 
( \tilde g_0-2\tilde g_1) < 1
}

For the diagonal RS solution in which $q$ vanishes,
$r = 1$, we find that both conditions \ATcond\ and \Longcond\ 
are the same and we recover 
the same line \Divcond\ for the spin glass transition.

\listrefs
%\listfigs
\bye